\documentclass[10pt, a4paper]{article}
\usepackage{amsmath, amssymb, amsthm, physics, siunitx}
\usepackage{geometry, fancyhdr, graphicx, booktabs, multirow}
\usepackage{hyperref, enumerate, enumitem}

\newcommand{\vect}[1]{\mathbf{#1}}

\begin{document}

\title{\textbf{Analytical Expression for Spherically Symmetric Photoacoustic Sources: A Unified General Solution}}
\author{
Shuang Li\textsuperscript{1} \and
Yibing Wang\textsuperscript{1} \and
Yu Zhang\textsuperscript{1} \and
Changhui Li\textsuperscript{1,2,}\thanks{Corresponding author: \texttt{chli@pku.edu.cn}}
}

\date{
\textsuperscript{1}Department of Biomedical Engineering, College of Future Technology, Peking University, Beijing, China\\
\textsuperscript{2}National Biomedical Imaging Center, Peking University, Beijing, China
}
\maketitle

\begin{abstract}
Here we present a comprehensive derivation of the analytical expression for the spatiotemporal acoustic pressure generated by photoacoustic sources with spherically symmetric initial pressure distributions. Starting from the fundamental photoacoustic wave equation, we derive a unified analytical solution applicable to arbitrary spherically symmetric initial distributions. Specific expressions are provided for several common distributions including uniform spherical sources, Gaussian distributions, exponential distributions, and power-law distributions. Far-field approximations are also discussed. The derived expressions provide valuable tools for photoacoustic imaging system design and signal analysis. We provide codes for ultrafast forward simulation using the general analytical spherically symmetric model, the implementation is available in the GitHub repository: \href{https://github.com/JaegerCQ/SlingBAG_Ultra}{https://github.com/JaegerCQ/SlingBAG\_Ultra}.
\end{abstract}

\section{Theoretical Background}
\subsection{Photoacoustic Wave Equation}
The pressure wave generated by the photoacoustic effect satisfies the following wave equation~\cite{wang2007biomedical,xu2005universal,xu2006photoacoustic}:
\begin{equation}
\left(\nabla^2 - \frac{1}{v_s^2}\frac{\partial^2}{\partial t^2}\right)p(\vect{r},t) = -\frac{\beta}{C_p}\frac{\partial H(\vect{r},t)}{\partial t}
\label{eq:wave_eq}
\end{equation}
where:
\begin{itemize}
    \item $p(\vect{r},t)$ is the acoustic pressure
    \item $v_s$ is the speed of sound
    \item $\beta$ is the thermal expansion coefficient
    \item $C_p$ is the specific heat capacity
    \item $H(\vect{r},t)$ is the heating function
\end{itemize}

For instantaneous heating, the initial conditions are:
\begin{equation}
p(\vect{r},0) = p_0(\vect{r}), \quad \frac{\partial p}{\partial t}(\vect{r},0) = 0
\label{eq:initial_cond}
\end{equation}

\subsection{Fundamental Integral Formula}
\begin{figure}[htbp]
    \centering
    \includegraphics[width=0.55\textwidth]{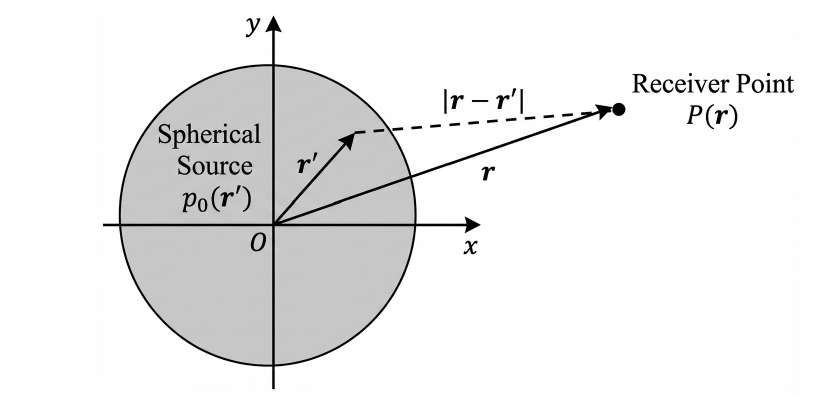}
    \caption{Schematic geometry of spherically symmetric photoacoustic wave propagation. 
    The photoacoustic source is located at the origin $O$, the field point is $P(\mathbf{r})$, 
    and $\mathbf{r}'$ denotes the position vector within the source region.}
    \label{fig:PA_setup}
\end{figure}

The solution to the photoacoustic wave equation can be expressed using Green's function as~\cite{wang2007biomedical,xu2005universal,xu2006photoacoustic}:
\begin{equation}
p(\vect{r},t) = \frac{1}{4\pi v_s^2}\frac{\partial}{\partial t}\left[\frac{1}{v_s t}\int d\vect{r}'\,p_0(\vect{r}')\,\delta\!\left(t-\frac{\abs{\vect{r}-\vect{r}'}}{v_s}\right)\right]
\label{eq:integral_form}
\end{equation}
where $p_0(\vect{r}')$ is the initial pressure distribution (Fig.~\ref{fig:PA_setup}).

\section{Derivation for Spherically Symmetric Case}
\subsection{Basic Assumptions}
Assume the initial pressure distribution is spherically symmetric about the origin:
\begin{equation}
p_0(\vect{r}') = p_0(r'), \quad r' = \abs{\vect{r}'}
\label{eq:symmetry}
\end{equation}
Denote $r = \abs{\vect{r}}$.

\subsection{Delta Function Transformation}
Using the scaling property of the delta function:
\begin{equation}
\delta\!\left(t-\frac{\abs{\vect{r}-\vect{r}'}}{v_s}\right) = v_s\,\delta\!\left(\abs{\vect{r}-\vect{r}'}-v_s t\right)
\label{eq:delta_transform1}
\end{equation}
Substituting into the original expression:
\begin{equation}
p(\vect{r},t) = \frac{1}{4\pi v_s^2}\frac{\partial}{\partial t}\left[\frac{1}{t}\int d\vect{r}'\,p_0(r')\,\delta\!\left(\abs{\vect{r}-\vect{r}'}-v_s t\right)\right]
\label{eq:pressure_intermediate}
\end{equation}

\subsection{Spatial Integral Treatment}
Consider the integral:
\begin{equation}
I = \int d\vect{r}'\,p_0(r')\,\delta\!\left(\abs{\vect{r}-\vect{r}'}-R\right)
\label{eq:integral_I}
\end{equation}
where $R = v_s t$.

Using spherical coordinates $(r',\theta,\phi)$, where $\theta$ is the angle between $\vect{r}'$ and $\vect{r}$. By the law of cosines:
\begin{equation}
\abs{\vect{r}-\vect{r}'} = \sqrt{r^2 + r'^2 - 2rr'\cos\theta}
\label{eq:cosine_law}
\end{equation}

After a series of transformations (detailed derivation in Appendix), we obtain:
\begin{equation}
\delta\!\left(\abs{\vect{r}-\vect{r}'}-R\right) = \frac{R}{rr'}\,\delta\!\left(\cos\theta - \frac{r^2 + r'^2 - R^2}{2rr'}\right)
\label{eq:delta_transform2}
\end{equation}

\subsection{Angular Integration}
Integrating over angles:
\begin{equation}
\int d\Omega\,\delta\!\left(\abs{\vect{r}-\vect{r}'}-R\right) = \frac{2\pi R}{rr'}
\label{eq:angular_integral}
\end{equation}
with the condition $\abs{\frac{r^2 + r'^2 - R^2}{2rr'}} \leq 1$, i.e., $\abs{r-R} \leq r' \leq r+R$.

Therefore:
\begin{equation}
I = \frac{2\pi R}{r}\int_{\abs{r-R}}^{r+R} r'p_0(r')\,dr'
\label{eq:I_result}
\end{equation}

\subsection{Time Derivative Calculation}
Substituting into the pressure expression:
\begin{equation}
p(r,t) = \frac{1}{4\pi v_s^2}\frac{\partial}{\partial t}\left[\frac{1}{t}\cdot\frac{2\pi v_s t}{r}\int_{\abs{r-v_s t}}^{r+v_s t} r'p_0(r')\,dr'\right]
\label{eq:pressure_before_simplify}
\end{equation}
Simplifying:
\begin{equation}
p(r,t) = \frac{1}{2v_s r}\frac{\partial}{\partial t}\left[\int_{\abs{r-v_s t}}^{r+v_s t} r'p_0(r')\,dr'\right]
\label{eq:pressure_simplified}
\end{equation}

Applying Leibniz's rule:
\begin{equation}
\frac{\partial}{\partial t}\int_{a(t)}^{b(t)} f(x)\,dx = f(b(t))\frac{db}{dt} - f(a(t))\frac{da}{dt}
\label{eq:leibniz}
\end{equation}
where $a(t) = \abs{r-v_s t}$, $b(t) = r+v_s t$.

\subsection{Final Analytical Expression}
After calculating the derivatives, we obtain the unified analytical expression for the spherically symmetric case:
\begin{equation}
\boxed{p(r,t) = \frac{1}{2r}\left[(r+v_s t)p_0(r+v_s t) + (r-v_s t)p_0(\abs{r-v_s t})\right]}
\label{eq:final_expression}
\end{equation}

\section{Analytical Solutions for Typical Initial Distributions}
\subsection{Uniform Spherical Source of Radius $a_0$}
Initial condition:
\begin{equation}
p_0(r) = p_0\,U(a_0 - r)
\label{eq:uniform_sphere}
\end{equation}
where $U(x)$ is the unit step function.

\subsubsection{Observation Point Outside the Sphere ($r > a_0$)}
\begin{itemize}
    \item When $v_s t < r - a_0$ or $v_s t > r + a_0$: $p(r,t) = 0$
    \item When $r - a_0 \leq v_s t \leq r + a_0$:
    \begin{equation}
    p(r,t) = \frac{p_0}{2r}(r - v_s t)
    \label{eq:uniform_outside}
    \end{equation}
\end{itemize}

\subsubsection{Observation Point Inside the Sphere ($r < a_0$)}
\begin{itemize}
    \item When $v_s t < a_0 - r$: $p(r,t) = p_0$
    \item When $a_0 - r \leq v_s t \leq a_0 + r$:
    \begin{equation}
    p(r,t) = \frac{p_0}{2r}(r - v_s t)
    \label{eq:uniform_inside}
    \end{equation}
    \item When $v_s t > a_0 + r$: $p(r,t) = 0$
\end{itemize}

\subsection{Gaussian Distribution}
Initial condition:
\begin{equation}
p_0(r) = p_c\,\exp\!\left(-\frac{r^2}{2\sigma^2}\right)
\label{eq:gaussian}
\end{equation}
Analytical solution:
\begin{equation}
p(r,t) = \frac{p_c}{2r}\left[(r+v_s t)\exp\!\left(-\frac{(r+v_s t)^2}{2\sigma^2}\right) + (r-v_s t)\exp\!\left(-\frac{(r-v_s t)^2}{2\sigma^2}\right)\right]
\label{eq:gaussian_solution}
\end{equation}

\subsection{Exponential Distribution}
Initial condition:
\begin{equation}
p_0(r) = p_c\,e^{-r/a}
\label{eq:exponential}
\end{equation}
Analytical solution:
\begin{equation}
p(r,t) = \frac{p_c}{2r}\left[(r+v_s t)e^{-(r+v_s t)/a} + (r-v_s t)e^{-\abs{r-v_s t}/a}\right]
\label{eq:exponential_solution}
\end{equation}

\subsection{Power-Law Distribution}
Initial condition:
\begin{equation}
p_0(r) = \frac{A}{(r^2 + a^2)^{\nu}}, \quad \nu > \frac{1}{2}
\label{eq:power_law}
\end{equation}
Analytical solution:
\begin{equation}
p(r,t) = \frac{A}{2r}\left[\frac{r+v_s t}{\big((r+v_s t)^2 + a^2\big)^{\nu}} + \frac{r-v_s t}{\big((r-v_s t)^2 + a^2\big)^{\nu}}\right]
\label{eq:power_law_solution}
\end{equation}

Common special case $\nu = \frac{3}{2}$:
\begin{equation}
p(r,t) = \frac{A}{2r}\left[\frac{r+v_s t}{\big((r+v_s t)^2 + a^2\big)^{3/2}} + \frac{r-v_s t}{\big((r-v_s t)^2 + a^2\big)^{3/2}}\right]
\label{eq:power_law_3_2}
\end{equation}

\section{Far-Field Approximations for Specific Distributions}
Under far-field conditions ($r \gg \text{characteristic source dimension}$ and $t \approx r/v_s$), the general expression \eqref{eq:final_expression} can be significantly simplified for specific initial distributions. In this regime, the first term $(r+v_s t)p_0(r+v_s t)$ represents an inward converging wave that is exponentially or power-law suppressed near the observation time window. The dominant contribution comes from the second term $(r-v_s t)p_0(\abs{r-v_s t})$, leading to the unified far-field approximation:
\begin{equation}
p(r,t) \approx \frac{1}{2r}(r-v_s t)p_0(\abs{r-v_s t})
\label{eq:far_field_general}
\end{equation}

\subsection{Far-Field Solution for Gaussian Distribution}
Applying the far-field condition \eqref{eq:far_field_general} to the Gaussian distribution \eqref{eq:gaussian}:
\begin{equation}
p(r,t) \approx \frac{p_c}{2r}(r-v_s t)\exp\!\left(-\frac{(r-v_s t)^2}{2\sigma^2}\right)
\label{eq:gaussian_far_field}
\end{equation}
This represents a Gaussian pulse propagating outward with speed $v_s$.

\subsection{Far-Field Solution for Exponential Distribution}
Applying the far-field condition \eqref{eq:far_field_general} to the exponential distribution \eqref{eq:exponential}:
\begin{equation}
p(r,t) \approx \frac{p_c}{2r}(r-v_s t)e^{-\abs{r-v_s t}/a}
\label{eq:exponential_far_field}
\end{equation}
This represents an exponentially decaying pulse propagating outward with speed $v_s$. The characteristic decay length in space is $a$.

\subsection{Far-Field Solution for Power-Law Distribution}
Applying the far-field condition \eqref{eq:far_field_general} to the power-law distribution \eqref{eq:power_law}:
\begin{equation}
p(r,t) \approx \frac{A}{2r}\frac{r-v_s t}{\big((r-v_s t)^2 + a^2\big)^{\nu}}
\label{eq:power_law_far_field}
\end{equation}
For the special case $\nu = 3/2$:
\begin{equation}
p(r,t) \approx \frac{A}{2r}\frac{r-v_s t}{\big((r-v_s t)^2 + a^2\big)^{3/2}}
\label{eq:power_law_3_2_far_field}
\end{equation}

\subsection{Far-Field Solution for Uniform Spherical Source}
For a uniform spherical source of radius $a_0$ with observation point outside the sphere ($r > a_0$), the exact solution \eqref{eq:uniform_outside} is already in a simple form. In the far-field limit $r \gg a_0$, and near the arrival time $t \approx r/v_s$, we have:
\begin{equation}
p(r,t) = \frac{p_0}{2r}(r - v_s t) \quad \text{for } r-a_0 \leq v_s t \leq r+a_0
\label{eq:uniform_far_field}
\end{equation}
This shows a linear dependence on $(r-v_s t)$ in the far field.

\appendix
\section{Detailed Derivations}
\subsection{Detailed Delta Function Transformation}
Starting from:
\begin{equation}
\delta\!\left(\sqrt{A} - R\right), \quad A = r^2 + r'^2 - 2rr'\cos\theta
\label{eq:appendix_A}
\end{equation}
Let $g(A) = \sqrt{A} - R$, then:
\begin{equation}
\frac{dg}{dA} = \frac{1}{2\sqrt{A}}
\label{eq:appendix_dg}
\end{equation}
At $A = R^2$:
\begin{equation}
\left|\frac{dg}{dA}\right| = \frac{1}{2R}
\label{eq:appendix_dg_at_R}
\end{equation}
Using the delta function transformation formula:
\begin{equation}
\delta(g(A)) = \frac{\delta(A - R^2)}{\abs{g'(R^2)}} = 2R\,\delta(A - R^2)
\label{eq:appendix_delta_transform}
\end{equation}
Thus:
\begin{equation}
\delta\!\left(\sqrt{A} - R\right) = 2R\,\delta(A - R^2)
\label{eq:appendix_final_delta}
\end{equation}

\subsection{Delta Function Transformation for $\cos\theta$}
Let:
\begin{equation}
h(\cos\theta) = r^2 + r'^2 - R^2 - 2rr'\cos\theta
\label{eq:appendix_h}
\end{equation}
The zero point is at:
\begin{equation}
\cos\theta_0 = \frac{r^2 + r'^2 - R^2}{2rr'}
\label{eq:appendix_cos0}
\end{equation}
The derivative is:
\begin{equation}
\abs{\frac{dh}{d(\cos\theta)}} = 2rr'
\label{eq:appendix_dh}
\end{equation}
Therefore:
\begin{equation}
\delta(h(\cos\theta)) = \frac{1}{2rr'}\,\delta\!\left(\cos\theta - \frac{r^2 + r'^2 - R^2}{2rr'}\right)
\label{eq:appendix_delta_cos}
\end{equation}

Combining with the previous factor $2R$, we obtain:
\begin{equation}
\delta\!\left(\abs{\vect{r}-\vect{r}'} - R\right) = \frac{R}{rr'}\,\delta\!\left(\cos\theta - \frac{r^2 + r'^2 - R^2}{2rr'}\right)
\label{eq:appendix_final_delta2}
\end{equation}

\section{Nomenclature}
\begin{tabular}{ll}
    $p(\vect{r},t)$ & Acoustic pressure at position $\vect{r}$ and time $t$ \\
    $p_0(\vect{r})$ & Initial pressure distribution \\
    $v_s$ & Speed of sound \\
    $\beta$ & Thermal expansion coefficient \\
    $C_p$ & Specific heat capacity \\
    $H(\vect{r},t)$ & Heating function \\
    $r$ & Radial distance from origin \\
    $t$ & Time \\
    $\delta(x)$ & Dirac delta function \\
    $U(x)$ & Unit step function \\
    $\sigma$ & Gaussian width parameter \\
    $a$ & Exponential decay length or power-law scale parameter \\
    $a_0$ & Radius of uniform sphere \\
\end{tabular}

\bibliographystyle{unsrt}   
\bibliography{references}  

@article{xu2005universal,
  title={Universal back-projection algorithm for photoacoustic computed tomography},
  author={Xu, Minghua and Wang, Lihong V},
  journal={Physical Review E—Statistical, Nonlinear, and Soft Matter Physics},
  volume={71},
  number={1},
  pages={016706},
  year={2005},
  publisher={APS}
}

@book{wang2007biomedical,
  title={Biomedical optics: principles and imaging},
  author={Wang, Lihong V and Wu, Hsin-i},
  year={2007},
  publisher={John Wiley \& Sons}
}

@article{xu2006photoacoustic,
  title={Photoacoustic imaging in biomedicine},
  author={Xu, Minghua and Wang, Lihong V},
  journal={Review of scientific instruments},
  volume={77},
  number={4},
  year={2006},
  publisher={AIP Publishing}
}
\end{document}